# Rotational dynamics, ionic conductivity, and glass formation in a ZnCl$_2$-based deep eutectic solvent

A. Schulz, P. Lunkenheimer[a], and A. Loidl

**AFFILIATIONS**

Experimental Physics V, Center for Electronic Correlations and Magnetism, University of Augsburg, 86135 Augsburg, Germany

[a]**Author to whom correspondence should be addressed:** peter.lunkenheimer@physik.uni-augsburg.de

**ABSTRACT**

Glass formation and reorientational motions are widespread, but often-neglected features of deep eutectic solvents, although both can be relevant for the technically important ionic conductivity at room temperature. Here we investigate these properties for two mixtures of ethylene glycol and ZnCl$_2$, which were recently considered as superior electrolyte materials for application in zinc-ion batteries. For this purpose, we employed dielectric spectroscopy performed in a broad temperature range, extending from the supercooled state at low temperatures up to the liquid phase around room temperature and beyond. We find evidence for a relaxation process arising from dipolar reorientation dynamics, which reveals the clear signatures of glassy freezing. This freezing also governs the temperature dependence of the ionic dc conductivity. We compare the obtained results with those for deep eutectic solvents that are formed by the same hydrogen-bond donor, ethylene glycol, but by two different salts, choline chloride and lithium triflate. The four materials reveal significantly different ionic and reorientational dynamics. Moreover, we find varying degrees of decoupling of rotational dipolar and translational ionic motions, which partly can be described by a fractional Debye-Stokes-Einstein relation. The typical glass-forming properties of these solvents strongly affect their room-temperature conductivity.

## I. INTRODUCTION

One of the most prominent possible applications of deep eutectic solvents (DESs) is their use as electrolytes in electrochemical devices like batteries, supercapacitors, or for electrodeposition.[1,2,3,4,5,6,7,8,9,10] Many of their properties make them superior alternatives to conventional electrolyte materials, e.g., they are less flammable, easier to produce, biocompatible, and sustainable.[2,3,9,11,12,13,14,15] Moreover, various DESs were found to meet or even exceed different benchmarks for electrochemical applications, e.g., a broad electrochemical window and high room-temperature conductivity.[1,2,3,4,5,6,7,8,9]

In DESs, the mixing of two or more components leads to a melting-point reduction, which makes them liquid around room temperature. Often, they consist of a molecular hydrogen-bond donor (HBD), e.g., ethylene glycol (EG) or urea, mixed with a salt that acts as a hydrogen-bond acceptor. To be useful as electrolytes in electrochemical applications, DESs should contain suitable ions, e.g., Li$^+$ for lithium-ion batteries. This can be either achieved by admixing small amounts of a corresponding salt to the DES[5] or by using such a salt as one of the main constituents of the DES.[16,17] In recent years, batteries involving zinc ions as charge carriers have attracted increasing interest because, compared to Li-ion batteries, they have many advantages like environmental friendliness, high safety, the abundance of Zn reserves, and relatively high energy density.[18,19] Zinc-based DESs were proposed as ideal electrolytes for such batteries.[7,8,11,20,21,22,23]

A very promising example is the mixture of ethylene glycol and ZnCl$_2$,[7,8,24] also considered as an electrolyte in supercapacitors,[25] which is investigated in the present work.

In general, electrolytes in electrochemical devices should have high ionic dc conductivity, $\sigma_{dc}$, exceeding $\sim 10^{-4}\,\Omega^{-1}\,cm^{-1}$ at room temperature. Understanding the ion dynamics on a microscopic level thus is desirable to develop better DESs electrolytes. Mixtures at the eutectic point usually are good glass formers. Indeed, upon cooling most DESs do not crystallize but instead become supercooled liquids before finally crossing over into an amorphous glass state.[12,26,27,28,29] While this happens well below room temperature, this glassy freezing also affects the room-temperature properties of DESs, including their conductivity.[27,28] It also leads to clear deviations of $\sigma_{dc}(T)$ from Arrhenius behavior.

Another often neglected fact is the presence of reorientational dipolar dynamics in most DESs. It arises because the HBDs and/or the added ion species often are asymmetric molecules revealing a dipolar moment. Moreover, supramolecular structures due to the aggregation of HBD molecules and salt ions can occur in DESs,[8,30,31,32,33] which can exhibit reorientational dynamics, as well. In general, reorientational motions in ionic conductors can be highly relevant for the ionic mobility. This was demonstrated for plastic crystals and ionic liquids and interpreted in terms of a "revolving-door" or "paddle-wheel" mechanism.[34,35,36,37,38] Interestingly, correlations of ionic dc conductivity and reorientational dynamics where recently also found for several DESs.[27,28,29,39,40,41]

Investigating such coupling effects and the glassy freezing in DESs is essential to achieve a better understanding of their ionic conductivity on a microscopic level and should help developing strategies for the optimization of their properties. Dielectric spectroscopy is ideally suited for such investigations: It is a well-established method for studying the continuous slowing down of molecular motions when approaching the glass transition and it is sensitive to both, translational motions of ions and reorientational motions of dipoles.[42,43] Nevertheless, related to the vast amount of literature on DESs, studies of these materials using dielectric spectroscopy are quite rare[26,27,28,29,39,40,41,44,45,46,47,48,49] and often their reorientational dynamics is disregarded.

In the present work, we provide dielectric-spectroscopy results on two $ZnCl_2$/EG mixtures with different molar ratios (1:4 and 1:2). The 1:2 composition ($ZnCl_2$/EG-1:2) can be directly compared to earlier dielectric results on ethaline, which is a 1:2 mixture of choline chloride (ChCl) with EG, i.e., it has the same HBD and anion but a larger and asymmetric cation. The 1:4 mixture ($ZnCl_2$/EG-1:4) corresponds to the eutectic composition of this DES[50] and was found to have higher ionic conductivity than three other investigated $ZnCl_2$/EG ratios, including $ZnCl_2$/EG-1:2.[8] We compare it to a 1:4 mixture of lithium triflate and EG (LiOTf/EG).[28] Notably, in the pioneering work by Abbott et al.[50], $ZnCl_2$/EG-1:4 was found to have the highest conductivity amongst four different $ZnCl_2$/HBD systems studied there.

## II. Experimental Details

$ZnCl_2$ was purchased from Thermo Scientific (99.999% purity) and EG from Alfa Aesar (99% purity) and both used as is. The two $ZnCl_2$/EG compositions were prepared by mixing appropriate amounts of their components inside a glass vessel at 80°C for several hours, leading to a homogeneous, translucent liquids. To prevent water uptake, the samples were kept in a dry nitrogen atmosphere during preparation and measurements. The dielectric measurements were performed with a frequency-response analyzer (Novocontrol Alpha-A analyzer). The liquids were filled into parallel-plate capacitors made of stainless steel with a diameter of 10 - 12 mm and plate distances between 1.0 and 1.2 mm. As noted in Ref. 27, the rather large plate distances lead to a shift of the electrode contributions in the dielectric response to lower frequencies and, thus, allow for a better deconvolution of the extrinsic and intrinsic contributions in the dielectric spectra.[51] For temperature regulation, a $N_2$-gas flow cryostat was used (Novocontrol Quatro). Additional differential scanning calorimetry (DSC) measurements were performed using a DSC 8500 from PerkinElmer with heating and cooling rates of 10 K/min.

## III. RESULTS AND DISCUSSION

### A. Dielectric spectra and glass transition

There are various ways of representing and evaluating the results of dielectric measurements of ionically conducting systems. Often such data are analyzed in terms of the dielectric modulus. However, as discussed, e.g., in Ref. 29, for systems with simultaneous reorientational and translational motions some problems may occur using this representation. Indeed, as already noted in the pioneering work on the modulus formalism,[52] it is intended for "ionic conductors which contain no permanent molecular dipoles". Therefore, in the present work we analyze the complex dielectric permittivity, $\varepsilon^* = \varepsilon' - i\varepsilon''$, which is well-established for materials with reorientational degrees of freedom.[42] In addition, we present the real part of the conductivity, related to $\varepsilon''$ via $\sigma' = \varepsilon_0 \varepsilon'' \omega$ (with $\varepsilon_0$ the permittivity of free space and $\omega = 2\pi\nu$) and enabling the direct determination of the dc conductivity.

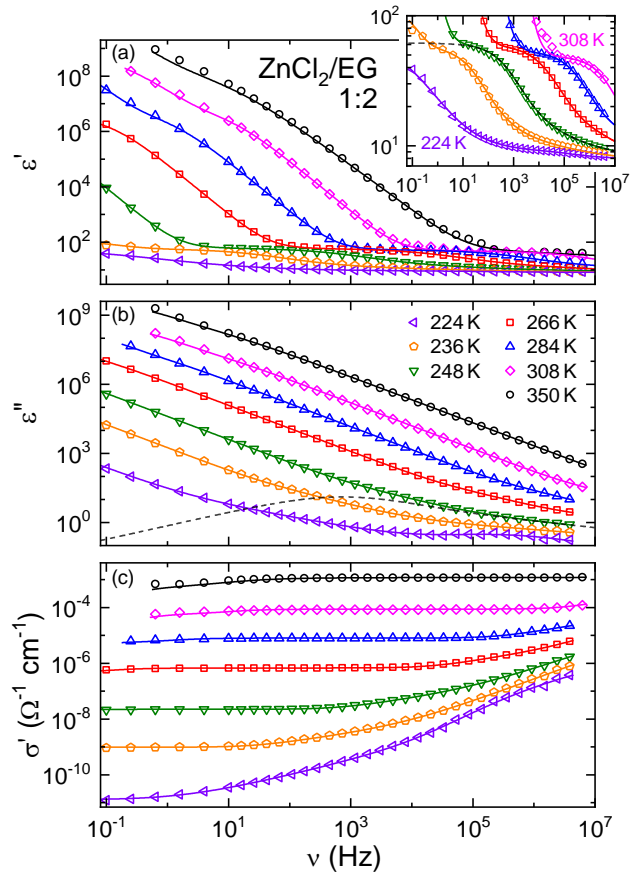

**FIG. 1.** Frequency dependence of the dielectric constant $\varepsilon'$ (a), dielectric loss $\varepsilon''$ (b), and real part of the conductivity $\sigma'$ (c) of $ZnCl_2$/EG-1:2, measured at various temperatures. The inset provides a zoomed-in view of the intrinsic relaxation revealed in $\varepsilon'(\nu)$. The solid lines in (a) and (b) are fits assuming an equivalent circuit including a distributed RC circuit to account for the electrode effects,[53] two intrinsic relaxation functions ($\alpha$ and secondary), and a dc-conductivity contribution as described in the text. The $\varepsilon'$ and $\varepsilon''$ spectra were simultaneously fitted for each temperature and the lines shown in (c) were calculated using $\sigma' = \varepsilon''\varepsilon_0 2\pi\nu$. As an example, the dashed lines in the inset and in (b) demonstrate the intrinsic contributions to the permittivity spectra at 248 K.

Figure 1 presents spectra of the dielectric constant ($\varepsilon'$), loss ($\varepsilon''$), and conductivity ($\sigma'$), measured at various temperatures for $ZnCl_2$/EG-1:2. For $ZnCl_2$/EG-1:4, we have obtained qualitatively similar results. The spectra resemble those reported for other DESs[27,28,29,39] and reveal the typical features of ionic conductors with additional reorientational dipole dynamics. In particular, at low frequencies and high temperatures, $\varepsilon'(\nu)$ [Fig. 1(a)] approaches unreasonably high values, beyond $10^8$ in the present case. This can be ascribed to electrode polarization arising at low frequencies, when the ions reach the sample electrodes.[53] At such low frequencies, their motion becomes impeded, which generates thin, poorly conducting space-charge regions that act as huge capacitors. This effect is also responsible for the low-frequency decrease of $\sigma'(\nu)$ observed at the highest temperatures [Fig. 1(c)]. Beyond the electrode-dominated regime, a step-like decrease with increasing frequency shows up in the $\varepsilon'$ spectra (see inset of Fig. 1 for a zoomed-in view). This indicates the presence of a relaxation process due to dipolar reorientational motions (termed $\alpha$ relaxation).[42,43] Upon cooling, these steps shift to lower frequencies by several decades, which is typical for the slowing down of the molecular dynamics found in dipolar glass-forming materials.[43,54] Indeed, DSC measurements reveal clear signatures of a glass transition for both compounds (Fig. 2).

relaxation process is detected.[55,56] However, we cannot exclude that slower relaxation processes may be hidden under the dominating contributions from dc conductivity (in $\varepsilon''$) or electrode polarization (in $\varepsilon'$). As pointed out in Ref. 57 and also applied to DESs,[41,58] plotting the derivative of $\varepsilon'$ may help to unravel such hidden processes. Figure 3 shows $-\partial\varepsilon'/\partial(\log\nu)$, which is approximately proportional to $\varepsilon''$,[57] for four temperatures where well-pronounced relaxation steps were detected. This quantity reveals a peak at the same position as the point of inflection in $\varepsilon'(\nu)$ (cf. inset of Fig. 1) as expected for dielectric-loss spectra. Interestingly, at lower frequencies a shoulder shows up which seems to indicate a second, slower and weaker relaxation process. It may well arise from reorientational motions of larger molecule/ion clusters in this DES. To unequivocally identify the microscopic origin of this spectral feature, e.g., nuclear-magnetic-resonance measurements would be desirable.

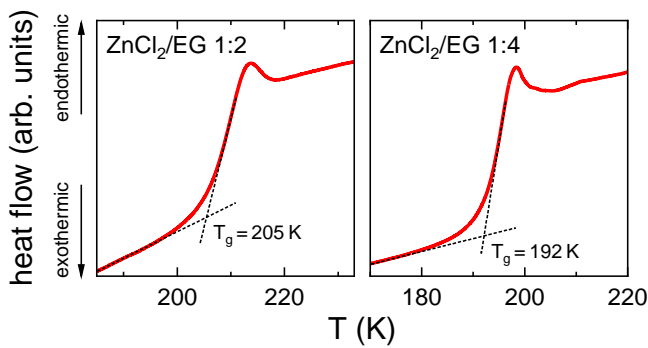

**FIG. 2.** DSC traces for the two DESs investigated in the present work, measured upon heating with 10 K/min. The glass-transition temperature $T_g$ was determined from the onset of the step-like increase in the heating trace, as indicated by the dashed lines.

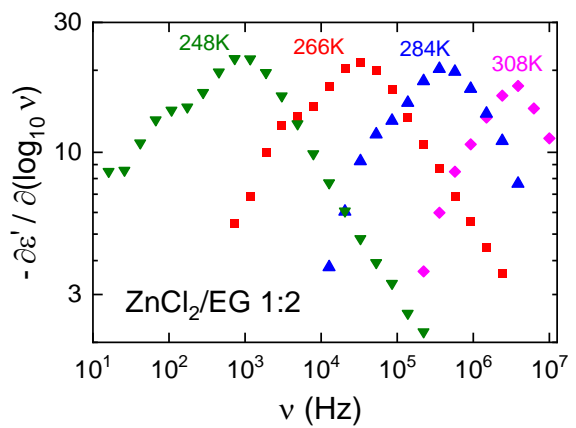

**FIG. 3.** Derivative of the $\varepsilon'$ spectra in Fig. 1(a) for four temperatures. The frequency range of each spectrum is restricted to the region where electrode polarization plays no role.

Aside from steps in $\varepsilon'(\nu)$, dipolar relaxation processes should also lead to peaks in the loss spectra. However, in Fig. 1(b) such peaks are not observed because the contribution from the dc-conductivity, $\varepsilon''_{dc} \propto \sigma_{dc}/\nu$, leads to a dominant $1/\nu$ increase at low frequencies. Therefore, only the right flanks of the expected peaks show up, which, at high frequencies, give rise to a shallower decrease of $\varepsilon''(\nu)$, compared to the dc contribution [cf. the dashed line in Fig. 1(b) indicating, for one temperature, the unobscured intrinsic dielectric response obtained from the fits described below].

In ZnCl$_2$/EG-1:2 and ZnCl$_2$/EG-1:4, reorientations of the dipolar EG molecules, which make up 66% or 80% of the sample constituents, respectively, can be assumed to play a major role in the generation of the detected relaxation process. However, the formation of other dipolar entities in ZnCl$_2$/EG mixtures, arising from complex aggregations between their ion and molecule constituents, was considered in several earlier works, e.g., [ZnCl(EG)]$^+$ cations[8,50] or ZnCl$_2$-4EG complexes.[33] We do not find any indications of separate relaxation processes from different dipolar entities in our data and probably all their rotations are closely coupled. Indeed, in materials with two different dipole species, often only a single

At the two lowest temperatures in Fig. 1(a), the relaxation steps in $\varepsilon'(\nu)$ occur below 1 kHz. However, in the corresponding loss spectra in Fig. 1(b), at high frequencies a region with only weak frequency dependence shows up, which indicates a further minor contribution in this region. It can be ascribed to a secondary relaxation process, usually termed $\beta$ relaxation, as often found in glass-forming liquids.[38,59,60,61,62] Similar additional high-frequency contributions were also reported for other DESs.[26,27,28,29] Their detailed treatment is out of the scope of the present work.

The ionic dc conductivity of ZnCl$_2$/EG-1:2 can be directly read off from the frequency-independent plateaus in $\sigma'(\nu)$ [Fig. 1(c)]. Its strong temperature-dependent variation by many decades arises from the essentially thermally-activated nature of the ionic mobility. The crossover of $\sigma'(\nu)$ from the dc plateau to a region with increasing conductivity at high frequencies is due to the above-discussed dipolar relaxation contributions detected in $\varepsilon''(\nu)$ because both quantities are related via $\sigma' \propto \varepsilon''\nu$. It should be noted that an increasing $\sigma'(\nu)$ at high frequencies often is also interpreted in terms of ac conductivity due to ionic hopping. Such an increase is predicted by a variety of competing models on ionic charge transport, e.g., the jump relaxation model[63] or the random free-energy barrier hopping model (RBM).[64,65] Indeed, often similar data on ionic conductors as shown in Fig. 1 are analyzed in terms of such models. However, one should be aware that an analysis of the apparent ac conductivity within the framework of such models only accounts for contributions from translational ion hopping and neglects any dipolar reorientation dynamics. On the other hand, the reorientational motions of the HBDs, making up significant parts of the investigated DESs, will inevitably lead to signatures of dipolar relaxation processes in the dielectric spectra. As discussed in more detail below, the present DES spectra can be perfectly fitted by solely assuming dipolar reorientation dynamics and dc charge transport, without any additional ac conductivity. Considering Occam's razor, in the following we therefore analyze our data without invoking ac conductivity, although we cannot fully exclude that such contributions may exist, superimposed by the dominating spectral features from dipolar dynamics. For a more detailed discussion of these issues and for examples of alternative fits of DES spectra including contributions predicted by the RBM, we refer the reader to our earlier works.[27,28,29]

To obtain reliable information on the intrinsic relaxation parameters, it is essential to fit the experimental dielectric spectra, including the mentioned electrode effects. To account for these non-intrinsic contributions, we assumed a distributed RC circuit that is connected in series to the bulk sample.[53] Such an equivalent-circuit approach was previously used for various ionic conductors, including DESs.[27,29,37,38] The intrinsic $\alpha$ and secondary relaxations were fitted by the sum of two Cole-Cole (CC) functions, an empirical fit function often used for disordered matter.[42,43,66,67] One should note that the CC function, leading to symmetric loss peaks, is well-established for secondary relaxations, but it is less commonly employed to describe the $\alpha$ relaxation, the latter often leading to asymmetric peaks. However, in the DES glyceline, the CC function was also found to provide good fits, just as in the present case.[29,68] In contrast, our qualitatively similar spectra for $ZnCl_2/EG$-1:4 could be somewhat better fitted assuming the asymmetric Cole-Davidson function[69] for the $\alpha$ relaxation. In Ref. 29, the superposition of the main relaxation peak by a low-frequency ac-conductivity contribution was considered as possible reason for the better description of the glyceline spectra by the CC function. However, as the conductivity in $ZnCl_2/EG$-1:4 is much higher than for the 1:2 composition (as discussed in the following section), it is unlikely that such a contribution is present for the 1:2 but absent for the 1:4 sample. In general, the broadening of the loss peaks described by these functions is ascribed to a distribution of relaxation times due to heterogeneity.[70,71] It seems reasonable that EG is the main dipolar constituent in these two systems and that the relaxation times of these molecular dipoles vary, depending on the disordered environment sensed by each molecule within the liquid. This environment is given by other EG molecules, the salt ions, and various clusters that may form in these DESs.[8,33,50] Our results indicate that the resulting distribution of relaxation times is qualitatively different for the significantly different concentrations of dipolar EG in the two compounds.

Finally, in the fits the dc-conductivity contribution to the loss was accounted for by a term $\varepsilon''_{dc} = \sigma_{dc}/(\varepsilon_0\omega)$. The solid lines in Fig. 1 are fits with these contributions, which were simultaneously performed for $\varepsilon'(\nu)$ and $\varepsilon''(\nu)$. The fit lines for $\sigma'(\nu)$ were calculated from those for $\varepsilon''(\nu)$. The fits provide a very good description of the experimental spectra. It should be noted that, at high temperatures, the secondary relaxation can be neglected as it is shifted out of the frequency window. In a similar way, at low temperatures the electrode effects play no role. Therefore, the number of fit parameters remains within reasonable limits.

### B. dc conductivity and $\alpha$-relaxation time

Figure 4(a) shows an Arrhenius representation of the temperature-dependent dc conductivities of $ZnCl_2/EG$-1:2 (closed circles) and $ZnCl_2/EG$-1:4 (closed squares) as deduced from the fits of the dielectric spectra. For both DESs, we find clearly non-linear behavior that evidences strong deviations from simple thermally-activated behavior given by $\sigma_{dc} \propto \exp(-E/[k_BT])$, where $E$ is an energy barrier and $k_B$ the Boltzmann constant. This is typical for glass-forming ionic conductors[72] and was also found for other DESs measured in a sufficiently broad temperature range.[5,6,26,27,28,29,39,68] It arises from the coupling of the ionic motion to the characteristic non-Arrhenius behavior revealed by the structural dynamics of glass-forming liquids.[43,54]

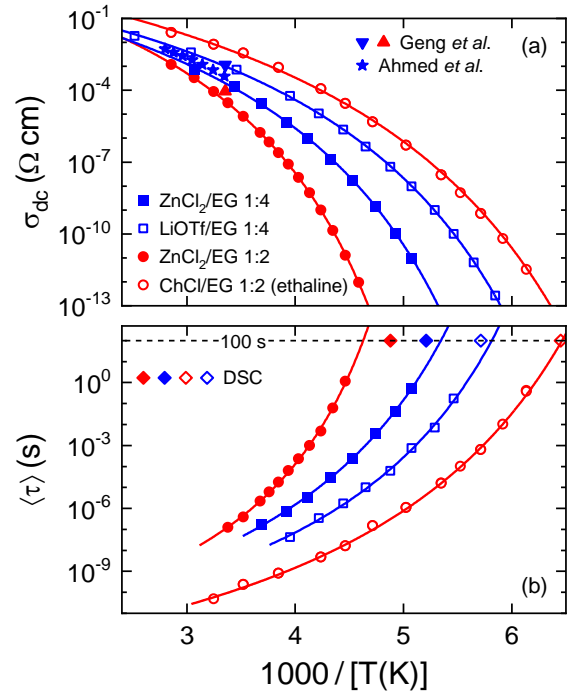

**FIG. 4.** Arrhenius plot of the dc conductivities (a) and mean $\alpha$-relaxation times (b) for $ZnCl_2/EG$-1:2 and $ZnCl_2/EG$-1:4 (present work) and for ethaline[27,29] and LiOTf/EG[28] (circles: DESs with 1:2 molar ratio, squares: 1:4). The upright and inverted triangles in (a) show $\sigma_{dc}$ at room temperature of $ZnCl_2/EG$-1:2 and $ZnCl_2/EG$-1:4, respectively, as reported in Ref. 8. The stars in (a) present $\sigma_{dc}(T)$ data for $ZnCl_2/EG$-1:4 from Ref. 24. Presuming $\langle\tau\rangle(T_g) \approx 100$ s, the diamonds in (b) indicate the glass-transition temperatures determined by DSC measurements (Fig. 2). The horizontal dashed line in (b) denotes $\langle\tau\rangle = 100$ s. The solid lines in (a) and (b) represent fits with the VFT laws, Eqs. (1) and (2), respectively.

At low temperatures, the conductivity of the eutectic mixture ZnCl$_2$/EG-1:4 exceeds that of ZnCl$_2$/EG-1:2 by many decades. With increasing temperature, the $\sigma_{dc}(T)$ traces of both mixtures approach each other, but at room temperature the conductivity of the eutectic composition still is by about one decade higher. Enhanced $\sigma_{dc}$ of the 1:4 mixture at room temperature was also found in Ref. 8. However, the absolute values reported there, shown by the triangles in Fig. 4(a) for both molar ratios, are somewhat higher than in the present work, which may indicate a higher water content of the investigated samples. In contrast, the $\sigma_{dc}(T)$ results for the 1:4 composition from Ref. 24, shown by the closed stars in Fig. 4(a), reasonably agree with the present data. This is also the case for the data in Ref. 50 (not shown). Finally, the dc conductivities reported in Ref. 73 for ZnCl$_2$/EG-1:2 around room temperature are up to one decade higher than the present results, pointing to a rather high water content of the samples investigated there.

As reported, e.g., in Ref. 29, the conductivity of DESs often is largely governed by their viscosity $\eta$, which seems plausible when considering the naive picture of charged particles, translationally moving within a viscous medium. Especially for ethaline, having the same HBD as in the present case, such coupling is well fulfilled.[29] In Ref. 8, the room-temperature viscosity of ZnCl$_2$/EG-1:4 was found to be significantly lower than for ZnCl$_2$/EG-1:2, which is in accord with its higher conductivity. The lower viscosity for the 1:4 compound is in line with the finding that, just for this eutectic composition, the ZnCl$_2$ admixture is most effective in breaking up the H-bonded network of pure EG.[33] Moreover, the infrared and Raman spectroscopy in Ref. 33 revealed that this eutectic composition exhibits the highest solvation of ZnCl$_2$, which also should contribute to an enhanced ionic conductivity of ZnCl$_2$/EG-1:4. As pointed out in Refs. 27 and 28, the glass-transition temperature $T_g$ is another important factor that can lead to strong deviations in the conductivity of different DESs, especially at low temperatures. The DSC results of Fig. 2 lead to $T_g$ = 192 and 205 K for the 1:4 and 1:2 compositions, respectively (Table I). As the viscosity at the glass transition is large (~10$^{12}$ Pa s), the higher $T_g$ of ZnCl$_2$/EG-1:2 causes a stronger increase of viscosity upon cooling and, thus, lower conductivity, as observed in Fig. 4(a).

As mentioned in section I, we compare these data to earlier results on other DESs with the same HBD. While, ethaline has an identical salt/HBD molar ratio as ZnCl$_2$/EG-1:2, its choline$^+$ cations are asymmetric and much larger than Zn$^{2+}$. In Fig. 4(a), the dc conductivity of ethaline is shown by the open circles.[27,29] At room temperature, it exceeds that of ZnCl$_2$/EG-1:2 by more than two decades and, at low temperatures, the difference becomes huge. At first glance, this may seem unexpected as the smaller zinc cations in the latter compound may be thought to reveal higher mobility. However, here again the different glass temperatures of the systems come into play, especially at low temperatures, with ethaline having a much smaller $T_g$ (155 K)[27] than ZnCl$_2$/EG-1:2 (205 K). The small size of the zinc ion and its twice as large charge lead to a higher charge-to-radius ratio (sometimes termed ionic potential) than for the ions in choline-chloride. This leads to stronger interactions, both inter-ionic and between the Zn$^{2+}$ ions and the HBD molecules. It can be expected to result in a stronger structural network within the liquid and, consequently, a higher viscosity and glass-transition temperature. Moreover, enhanced interactions of the zinc ions with the other constituents and among each other should also directly reduce their mobility. In general, a reduction of conductivity when replacing larger by smaller ions in an ionic conductor is an often-found phenomenon and was, e.g., also reported for ionic liquids.[74,75]

The open squares in Fig. 4(a) show $\sigma_{dc}(T)$ of LiOTf/EG, which has the same salt/HBD ratio as ZnCl$_2$/EG-1:4. In this case, the OTf (triflate) anion is larger than the Cl$^-$ anions and asymmetric while Zn$^{2+}$ and Li$^+$ have comparable sizes. Again, the conductivity is significantly higher and the $T_g$ is lower than for the corresponding ZnCl$_2$ system, but the differences are less extreme than for the 1:2 compositions. It is clear that all these conductivity variations should also depend on details like different cluster formation between the DES constituents,[8,33,50] but the general relevance of the glass transition and the increased interactions due to the smaller ions in the ZnCl$_2$-based DESs is confirmed by these results.

**TABLE I.** Glass-transition temperatures deduced by DSC (Fig. 2) and VFT parameters from the fits of $\sigma_{dc}(T)$ and $\langle\tau\rangle(T)$ shown in Fig. 4. The fragility parameter $m$ was calculated from $D_\tau$.[81] For comparison, corresponding data are provided for ethaline[27] and LiOTf/EG.[28]

| | $T_g$ (K) | $T_{VF\sigma}$ (K) | $D_\sigma$ | $\sigma_0$ ($\Omega^{-1}$cm$^{-1}$) | $T_{VF\tau}$ (K) | $D_\tau$ | $\tau_0$ (s) | $m$ |
|---|---|---|---|---|---|---|---|---|
| ZnCl$_2$/EG-1:2 | 205 | 156 | 12.2 | 24 | 168 | 9.40 | 5.6×10$^{-13}$ | 79 |
| ZnCl$_2$/EG-1:4 | 192 | 135 | 12.6 | 6.3 | 135 | 12.9 | 4.6×10$^{-13}$ | 62 |
| ethaline (ChCl/EG 1:2) | 155 | 111 | 13.7 | 81 | 113 | 13.2 | 2.8×10$^{-14}$ | 60 |
| LiOTf/EG (1:4) | 175 | 122 | 12.4 | 5.3 | 128 | 11.4 | 4.8×10$^{-13}$ | 68 |

The $\sigma_{dc}(T)$ data of Fig. 4(a) can be well fitted by a modification of the empirical Vogel-Fulcher-Tammann (VFT) law[76,77,78] [lines in Fig. 4(a)]:

$$\sigma_{dc} = \sigma_0 \exp\left[\frac{-D_\sigma T_{VF\sigma}}{T - T_{VF\sigma}}\right] \quad (1)$$

Here $\sigma_0$ is a pre-exponential factor, $D_\sigma$ is the so-called strength parameter,[79] and $T_{VF\sigma}$ is the Vogel-Fulcher temperature. The resulting fit parameters are listed in Table I. The VFT law is a characteristic feature of glass-forming matter, mostly applied to viscosity and relaxation-time data.[43,54,80] It describes the typical super-Arrhenius slowing down of the dynamics when approaching the glass temperature, where $D_\sigma$ quantifies the deviations from Arrhenius behavior.[79] It was previously also applied to $\sigma_{dc}(T)$ data of other DESs.[5,6,24,27,28,29]

The average reorientational $\alpha$-relaxation times $\langle\tau\rangle$, obtained from the fits of the dielectric spectra of the present ZnCl$_2$/EG DESs, are shown in Fig. 4(b) (closed symbols), using an Arrhenius representation. Corresponding data for ethaline and LiOTf/EG are indicated by the open symbols.[27,28,29] Similar to the dc conductivity, which is a measure of the translational ion dynamics [Fig. 4(a)], the relaxation times quantifying the dipolar rotation dynamics reveal significant deviations from Arrhenius behavior, characteristic of glass-forming liquids.[43,54,80] Interestingly, the dc-conductivity curves in Fig. 4(a) appear like approximate mirror images of the relaxation-time traces in Fig. 4(b), indicating that the translational ionic and the reorientational dipolar dynamics are coupled to some extent. This finding is in accord with the viscosity-related arguments invoked in the discussion of the conductivity results above, when considering the simple picture of asymmetric particles rotating within a viscous medium. Indeed, the translation-rotation coupling previously evidenced in several DESs, including ethaline,[27] was found to arise from the strong link of both dynamics to the viscosity.[29]

The solid lines in Fig. 4(b) show fits of $\langle\tau\rangle(T)$ by the VFT law, Eq. (2).[76,77,78,79]

$$\langle\tau\rangle = \tau_0 \exp\left[\frac{D_\tau T_{VFT\tau}}{T - T_{VFT\tau}}\right] \quad (2)$$

The obtained parameters are listed in Table I. For ZnCl$_2$/EG-1:4, the Vogel-Fulcher temperatures and strength parameters from the fits of $\langle\tau\rangle(T)$ and $\sigma_{dc}(T)$ are very similar, which also indicates significant coupling of the reorientational and ionic dynamics. This resembles the behavior in ethaline, where such coupling was demonstrated in Refs. 27 and 29. In contrast, the deviations revealed in Table I for the VFT parameters of ZnCl$_2$/EG-1:2 points to less perfect coupling as will be treated in more detail in the next section.

From $\langle\tau\rangle(T)$ data, the glass-transition temperature can be estimated using the criterion $\langle\tau\rangle(T_g) \approx 100$ s. An extrapolation of the VFT fit curves in Fig. 4(b) leads to 216 and 187 K for ZnCl$_2$/EG-1:2 and ZnCl$_2$/EG-1:4, respectively. As noted in Table I, the DSC experiments on these systems result in $T_g \approx 205$ and 192 K, respectively. These temperatures are indicated by the diamonds at $\langle\tau\rangle = 100$ s in Fig. 4(b). The somewhat different glass-transition temperatures from the two methods could indicate some decoupling of the reorientational dynamics from the "true" glass transition at low temperatures, assumed to be detected by the entropy-sensitive calorimetry.

The so-called fragility parameter $m$, which can be calculated from $D_\tau$ via $m \approx 16 + 590/D_\tau$,[81] is the most common quantity for the characterization of the deviations of $\langle\tau\rangle(T)$ from simple thermally activated behavior. We deduce values of 79 and 62 for ZnCl$_2$/EG-1:2 and ZnCl$_2$/EG-1:4, respectively (Table I.). Thus, these systems can be considered as intermediate within the strong-fragile classification of glass-forming liquids[79], just as the other two DESs containing EG.[27,28] In principle, the fragility of a glass-forming liquid affects its room-temperature conductivity which was explicitly demonstrated for ionic liquids in Ref. 82. In the present case, the $m$ values of the four systems in Fig. 4 do not vary dramatically, and, thus, the significant differences in their conductivity cannot be ascribed to their fragilities.

## C. Coupling of reorientational and translational dynamics

To directly check the coupling of the translational ion and dipolar reorientation dynamics, Fig. 5 shows, for both compositions, the temperature dependences of $\rho_{dc} = 1/\sigma_{dc}$ and $\langle\tau\rangle$ within the same frame.[83] The two corresponding ordinates (left: $\langle\tau\rangle$; right: $\rho_{dc}$) are scaled to display the same number of decades for both quantities. Moreover, we adjusted the starting values of the y-axes to reach an agreement at high temperatures, since there decoupling effects are usually less important. While for ZnCl$_2$/EG-1:2 [Fig. 5(a)] some small but significant deviations show up at low temperatures, ZnCl$_2$/EG-1:4 [Fig. 5(b)] reveals a perfect match of the two curves, indicating direct proportionality of $\rho_{dc}$ and $\langle\tau\rangle$. Such perfect coupling was also found for ethaline[27,29] while the behavior of LiOTf/EG resembles that of ZnCl$_2$/EG-1:2.[28] As the Li-salt system has 1:4 and ethaline 1:2 molar ratio, it becomes clear that the EG content is not the sole factor determining the decoupling. Especially, when trying to understand the different coupling in ZnCl$_2$/EG-1:2 and ZnCl$_2$/EG-1:4, one should consider that, for the latter, the ZnCl$_2$ admixture was found to be most effective in breaking up the H-bonded network of pure EG.[33] Moreover, the highest solvation of ZnCl$_2$ was found for this 1:4 eutectic composition.[33] Therefore, it seems that a large number of dissolved ions moving within a liquid without extensive hydrogen network favors the coupling.

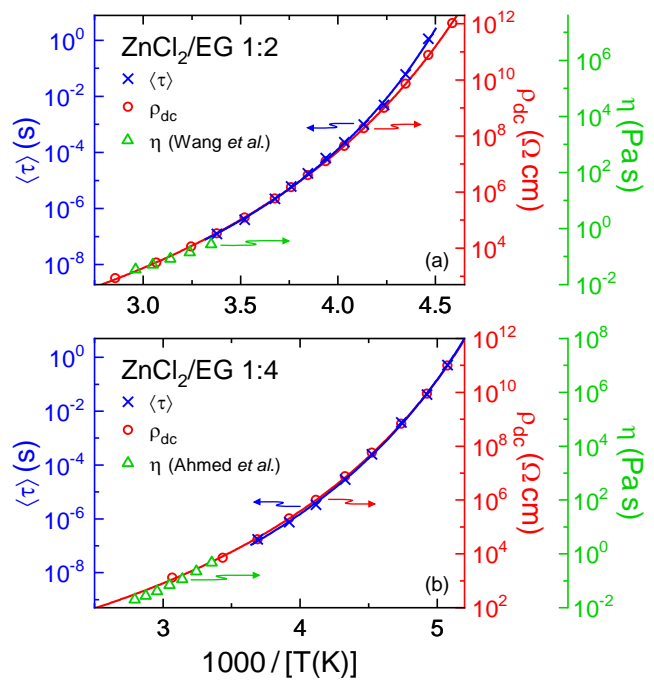

**FIG. 5.** Temperature dependences of the average $\alpha$-relaxation times (crosses, left ordinates) and dc resistivities (circles, right ordinates) of ZnCl$_2$/EG-1:2 (a) and ZnCl$_2$/EG-1:4 (b), plotted using an Arrhenius representation. The number of decades covered by the left and right y-axes in each frame is identical. The starting values of the ordinates were chosen to reach a match between both quantities at the highest investigated common temperature. The lines are the VFT fits from Fig. 4 [Eqs. (1) and (2)], taking into account $\rho_{dc} = 1/\sigma_{dc}$. The triangles in (a) and (b) show viscosity data from Refs. 73 and 24, respectively (rightmost ordinate), scaled to match the other data sets.

It should be noted that three ChCl-based DESs, including ethaline, have recently been found to reveal nearly perfect reorientation-viscosity coupling.[29] Unfortunately, to our knowledge, temperature-dependent viscosity data for the present ZnCl$_2$/EG DESs have only been reported in a rather limited temperature range.[24,50] The triangles depicted in Figs.

5(a) and (b) show $\eta(T)$ data from Refs. 73 and 24, respectively, which are scaled to match the high-temperature results of the other quantities. One should note that those from Ref. 73 may be hampered by rather high water content of the investigated samples, which can be concluded from the much higher dc conductivity reported in that work [e.g., $3.6 \times 10^{-4}$ $\Omega^{-1}$cm$^{-1}$ instead of $3.5 \times 10^{-5}$ $\Omega^{-1}$cm$^{-1}$ at 298 K deduced in the present work, cf. Fig. 4(a)]. Overall, the available viscosity data do not allow for any clear conclusions on possible reorientation-viscosity coupling as found in Ref. 29. When, nevertheless, assuming that such coupling also exists for the present systems, the reorientational relaxation times in Fig. 5 should have the same temperature dependence as the viscosity. The deviations seen in Fig. 5(a) then imply that the simple picture of a sphere, translationally moving within a viscous medium, is not valid at low temperatures in ZnCl$_2$/EG-1:2. It thus seems that the fewer free ions in the 1:2 system (due to less solvation of ZnCl$_2$, compared to 1:4)[33] find ways to diffuse faster than expected within the more extended H-bonded network governing the high viscosity of this composition,[8] which enhances its low-temperature conductivity.

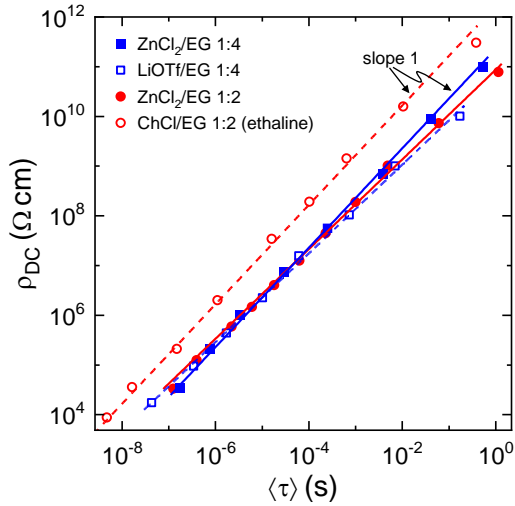

**FIG. 6.** Double-logarithmic plot of the dc resistivity vs the mean reorientational relaxation time for the two investigated DESs and the two reference systems.[27,28,29] The lines are linear fits with slope $\xi$, corresponding to power laws $\rho_{dc} \propto \langle \tau \rangle^{\xi}$. The data for ethaline and ZnCl$_2$/EG-1:4 can be well described with $\xi = 1$, implying $\rho_{dc} \propto \langle \tau \rangle$. For ZnCl$_2$/EG-1:2 and LiOTf/EG we obtain $\xi = 0.90$ and $\xi = 0.89$,[28] respectively.

Finally, Fig. 6 shows a double-logarithmic plot of the dc resistivity vs the $\alpha$-relaxation time for the two investigated DESs and reference systems. Just as previously reported for ethaline (open circles),[27] we find $\rho_{dc} \propto \langle \tau \rangle$ for ZnCl$_2$/EG-1:4 (closed squares), which is in accord with the Debye-Stokes-Einstein (DSE) relation.[84,85,86,87] This confirms the coupling of the ionic translational and dipolar reorientational dynamics revealed in Fig. 5(b). Interestingly, ZnCl$_2$/EG-1:2 (closed circles) exhibits a fractional power law, $\rho_{dc} \propto \langle \tau \rangle^{\xi}$ with $\xi = 0.90$, resembling the fractional DSE behavior detected in certain glass formers.[84,85] Similar behavior with $\xi = 0.89$ is found for LiOTf/EG (open squares in Fig. 6).[28] For given relaxation times, Fig. 6 reveals that the three DESs with small cations have clearly higher dc conductivity (smaller dc resistivity) than ethaline. When assuming the mentioned reorientation-viscosity coupling,[29] this implies that the smaller ions can move faster for a given viscosity which seems plausible. However, one should be aware that for a given *temperature* the conductivity of ethaline is the highest among these four DESs [see Fig. 4(a)]. This is due to the higher glass-transition temperatures (and, thus, viscosities) of the DESs with small cations. As discussed above, this can be ascribed to the stronger structural network of these liquids, due to their smaller ions and the higher charge of the zinc ions. Notably, without the decoupling evidenced by the fractional DSE relation, $\sigma_{dc}$ of these systems would be even lower at low temperatures.

## IV. SUMMARY AND CONCLUSIONS

In summary, we have performed dielectric spectroscopy on ZnCl$_2$-based DESs with two different salt-HBD molar ratios. Our measurements cover a temperature range extending from the low-viscosity liquid, above room temperature, well down to the deeply supercooled state. In addition to electrode-polarization effects, we find clear evidence of an intrinsic relaxation process. It can be ascribed to reorientational dipolar motions, most likely dominated by the EG molecules making up 66% - 80% of the sample constituents. Upon cooling, this dipolar dynamics exhibits the characteristics of glassy freezing, namely heterogeneity, which gives rise to a distribution of relaxation times, and a clear non-Arrhenius temperature dependence of the latter. The relaxation-times of the two investigated compositions and of two other DESs, where the HBD is EG, vary significantly, which is due to their different glass-transition temperatures and viscosities. These differences most likely can be ascribed to variations in the ionic interaction strength and the effectiveness of breaking the H-bonded network. The temperature-dependent dc conductivities of these DESs also exhibit clear non-Arrhenius behavior, closely resembling that of the relaxation times. Therefore, the large differences in conductivity between these DESs are also mainly caused by their different glass-transition temperatures and interactions on a molecular level.

For ZnCl$_2$/EG-1:2, a detailed comparison of the temperature-dependent dc resistivities and relaxation times reveals some significant decoupling at low temperatures. Just as previously reported for some other DESs,[27,28] the 1:2 molar ratio exhibits a fractional DSE relation. In this mixture, the ions seem to explore paths within the liquid structure that allow for diffusion that is enhanced, compared to the expectations for a viscous medium at low temperatures. However, one should be aware that this enhancement is limited and does not play any role for the eutectic composition ZnCl$_2$/EG-1:4, which reveals perfect coupling and much higher conductivity. This DES represents a canonical liquid electrolyte in which the translational ion and reorientational dipole dynamics are both most likely governed by the viscosity and thus coupled. Its ionic conductivity, however, is significantly lower than that of ethaline, having the same HBD but larger cations [Fig. 4(a)].

Considering the glass-forming properties of these DESs, the room-temperature ionic conductivity of $ZnCl_2$/EG-1:4 could be enhanced by decreasing its glass-transition temperature, which may be achieved by admixing compounds that further break the hydrogen-bonded network, however without replacing the hydrogen bonds by stronger ionic ones. Increasing the fragility would be an alternative approach, as this would lead to a stronger curvature of the conductivity curve in Fig. 4(a), as discussed in detail in Ref. 82. This may be achieved by the admixture of additional components, keeping in mind that a more complex energy landscape caused by such admixing should enhance fragility.[88,89] The enhanced conductivities of ternary DES systems, based on $ZnCl_2$/EG-1:4, reported in Ref. 24, may well be explainable in light of these considerations, but measurements in a broader temperature range would be desirable to obtain reliable values of the fragility and glass-transition temperatures.

## ACKNOWLEDGMENTS

This work was supported by the Deutsche Forschungsgemeinschaft (grant No. LU 656/5-1) under Project No. 444797029. We thank Catharina Binder for performing part of the dielectric measurements.

## AUTHOR DECLARATIONS

### Conflict of Interest

The authors have no conflicts to disclose.

## DATA AVAILABILITY

The data that support the findings of this study are available from the corresponding author upon reasonable request.

---

## REFERENCES

[1] C. A. Nkuku and R. J. LeSuer, J. Phys. Chem. B, **111**, 13271 (2007).
[2] Q. Zhang, K. D. O. Vigier, S. Royer, and F. Jérôme, Chem. Soc. Rev. **41** 7108 (2012).
[3] E. L. Smith, A. P. Abbott, and K. S. Ryder, Chem. Rev. **114**, 11060 (2014).
[4] M. H. Chakrabarti, F. S. Mjalli, I. M. AlNashef, M. A. Hashim, M. A. Hussain, L. Bahadori, and C. T. J. Low, Renewable Sustainable Energy Rev. **30**, 254 (2014).
[5] L. Millia, V. Dall'Asta, C. Ferrara, V. Berbenni, E. Quartarone, F. M. Perna, V. Capriati, and P. Mustarelli, Solid State Ionics **323**, 44 (2018).
[6] O. E. Geiculescu, D. D. DesMarteau, S. E. Creager, O. Haik, D. Hirshberg, Y. Shilina, E. Zinigrad, M. D. Levi, D. Aurbach, and I. C. Halalay, J. Power Sources **307**, 519 (2016).
[7] G. M. Thorat, V.-C. Ho, and J. Mun, Front. Chem. **9**, 825807 (2021).
[8] L. Geng, J. Meng, X. Wang, C. Han, K. Han, Z. Xiao, M. Huang, P. Xu, L. Zhang, L. Zhou, and L. Mai, Angew. Chem. Int. Ed. **61**, e202206717 (2022).
[9] B. B. Hansen, S. Spittle, B. Chen, D. Poe, Y. Zhang, J. M. Klein, A. Horton, L. Adhikari, T. Zelovich, B. W. Doherty, B. Gurkan, E. J. Maginn, A. Ragauskas, M. Dadmun, T. A. Zawodzinski, G. A. Baker, M. E. Tuckerman, R. F. Savinell, and J. R. Sangoro, Chem. Rev. **121**, 123 (2021).
[10] M. E. Di Pietro and A. Mele, J. Mol. Liq. **338**, 116597 (2021).
[11] A. P. Abbott, G. Capper, D. L. Davies, R. K. Rasheed, and V. Tambyrajah, Chem. Comm. **2003**, 70 (2003).
[12] Y. Dai, J. van Spronsen, G.-J. Witkamp, R. Verpoorte, and Y. H. Choi, Anal. Chim. Acta **766**, 61 (2013).
[13] A. Paiva, R. Craveiro, I. Aroso, M. Martins, R. L. Reis, and A. R. C. Duarte, ACS Sustainable Chem. Eng. **2**, 1063 (2014).
[14] C. J. Clarke, W. C. Tu, O. Levers, A. Bröhl, and J. P. Hallett, Chem. Rev. **118**, 747 (2018).
[15] A. M. Navarro-Suárez and P. Johansson, J. Electrochem. Soc. **167**, 070511 (2020).
[16] V. Lesch, A. Heuer, B. R. Rad, M. Winterac, and J. Smiatek, Phys. Chem. Chem. Phys. **18**, 28403 (2016).
[17] H. Cruz, N. Jordao, A. L. Pinto, M. Dionisio, L. A. Neves, and L. C. Branco, ACS Sustainable Chem. Eng. **8**, 10653 (2020).
[18] C. Xu, B. Li, H. Du, F. Kang, Angew. Chem. Int. Ed. **51**, 933 (2012).
[19] L. Ma, M. Schroeder, O. Borodin, T. Pollard, M. Ding, C. Wang, and K. Xu, Nat. Energy **5**, 743 (2020).
[20] F. S. Ghareh Bagh, K. Shahbaz, F. S. Mjalli, M. A. Hashim, and I. M. AlNashef, J. Mol. Liq. **204**, 76 (2015).
[21] A. R. Mainar, E. Iruin, L. C. Colmenares, A. Kvasha, I. de Meatza, M. Bengoechea, O. Leonet, I. Boyano, Zh. Zhang, and J.A. Blazquez, J. Energy Storage **15**, 304 (2018).
[22] Y. Lv, Y. Xiao, L. Ma, C. Zhi, and S. Chen, Adv. Mater. **34**, 2106409 (2021).
[23] X. Lin, G. Zhou, M. J. Robson, J. Yu, S. C. T. Kwok, and F. Ciucci, Adv. Funct. Mater. **32**, 2109322 (2022).
[24] F. Ahmed, D. Paterno, G. Singh, and S. Suarez, Electrochimica Acta **462**, 142784 (2023).
[25] H. Tian, R. Cheng, L. Zhang, Q. Q. Fang, P. Ma, Y. Lv, and F. Wei, Mater. Lett. **301**, 130237 (2021).
[26] S. N. Tripathy, Z. Wojnarowska, J. Knapik, H. Shirota, R. Biswas, and M. Paluch, J. Chem. Phys. **142**, 184504 (2015).
[27] D. Reuter, C. Binder, P. Lunkenheimer, and A. Loidl, Phys. Chem. Chem. Phys. **21**, 6801 (2019).
[28] A. Schulz, P. Lunkenheimer, and A. Loidl, J. Chem. Phys. **155**, 044503 (2021).
[29] D. Reuter, P. Münzner, C. Gainaru, P. Lunkenheimer, A. Loidl, and R. Böhmer, J. Chem. Phys. **154**, 154501 (2021).
[30] D. V. Wagle, G. A. Baker, and E. Mamontov, J. Phys. Chem. Lett. **6**, 2924 (2015).
[31] C. D'Agostino, R. C. Harris, A. P. Abbott, L. F. Gladden, and M. D. Mantle, Phys. Chem. Chem. Phys. **13**, 21383 (2011).
[32] Y. Zhang, D. Poe, L. Heroux, H. Squire, B. W. Doherty, Z. Long, M. Dadmun, B. Gurkan, M. E. Tuckerman, and E. J. Maginn, J. Phys. Chem. B **124**, 5251 (2020).
[33] P. Kalhor, K. Ghandi, H. Ashraf, Z. Yu, Phys. Chem. Chem. Phys. **23**, 13136 (2021).
[34] L. Börjesson and L. M. Torell, Phys. Rev. B **32**, 2471 (1985).

35 E. I. Cooper and C. A. Angell, Solid State Ionics **18-19**, 570 (1986).
36 D. R. MacFarlane and M. Forsyth, Adv. Mater. **13**, 957 (2001).
37 K. Geirhos, P. Lunkenheimer, M. Michl, D. Reuter, and A. Loidl, J. Chem. Phys. **143**, 081101 (2015).
38 P. Sippel, S. Krohns, D. Reuter, P. Lunkenheimer, and A. Loidl, Phys. Rev. E **98**, 052605 (2018).
39 A. Jani, B. Malfait, and D. Morineau, J. Chem. Phys. **154**, 164508 (2021).
40 C. D'Hondt and D. Morineau, J. Mol. Liq. **365**, 120145 (2022).
41 S. Spittle, D. Poe, B. Doherty, C. Kolodziej, L. Heroux, M. A. Haque, H. Squire, T. Cosby, Y. Zhang, C. Fraenza, S. Bhattacharyya, M. Tyagi, J. Peng, R. A. Elgammal, T. Zawodzinski, M. Tuckerman, S. Greenbaum, B. Gurkan, C. Burda, M. Dadmun, E. J. Maginn, and J. Sangoro, Nat. Commun. **13**, 219 (2022).
42 *Broadband Dielectric Spectroscopy*, edited by F. Kremer and A. Schönhals (Springer, Berlin, 2002).
43 P. Lunkenheimer and A. Loidl, in *The Scaling of Relaxation Processes*, edited by F. Kremer and A. Loidl (Springer, Cham, 2018), p. 23.
44 P. J. Griffin, T. Cosby, A. P. Holt, R. S. Benson, and J. R. Sangoro, J. Phys. Chem. B **118**, 9378 (2014).
45 K. Mukherjee, A. Das, S. Choudhury, A. Barman, and R. Biswas, J. Phys. Chem. B **119**, 8063 (2015).
46 K. Mukherjee, E. Tarif, A. Barman, and R. Biswas, Fluid Phase. Equilib. **448**, 22 (2017).
47 K. Mukherjee, S. Das, E. Tarif, A. Barman, and R. Biswas, J. Chem. Phys. **149**, 124501 (2018).
48 H. Cruz, N. Jordão, P. Amorim, M. Dionísio, and L. C. Branco, ACS Sustainable Chem. Eng. **6**, 2240 (2018).
49 V. Agieienko and R. Buchner, Phys. Chem. Chem. Phys. **22**, 20466 (2020).
50 A. P. Abbott, J. C. Barron, K. S. Ryder, and D. Wilson, Chem. Eur. J. 13, 6495 (2007).
51 P. Lunkenheimer, S. Krohns, S. Riegg, S. G. Ebbinghaus, A. Reller and A. Loidl, Eur. Phys. J. Special Topics **180**, 61 (2010).
52 P. B. Macedo, C. T. Moynihan, and R. Bose, Phys. Chem. Glasses **13**, 171 (1972).
53 S. Emmert, M. Wolf, R. Gulich, S. Krohns. P. Lunkenheimer, and A. Loidl, Eur. Phys. J. B **83**, 157 (2011).
54 P. Lunkenheimer, U. Schneider, R. Brand, and A. Loidl, Contemp. Phys. **41**, 15 (2000).
55 K. Duvvuri and R. Richert, J. Phys. Chem. B 1**08**, 10451 (2004).
56 L.-M. Wang, Y. Tian, R. Liu, and R. Richert, J. Phys. Chem. B **114**, 3618 (2010).
57 M. Wübbenhorst and J. van Turnhout, J. Non-Cryst. Solids **305**, 40 (2002).
58 I. Alfurayj, C. C. Fraenza, Y. Zhang, R. Pandian, S. Spittle, B. Hansen, W. Dean, B. Gurkan, R. Savinell, S. Greenbaum, E. Maginn, and J. Sangoro, J. Phys. Chem. B **125**, 8888 (2021).
59 G. P. Johari and M. Goldstein, J. Chem. Phys. **53**, 2372 (1970).
60 A. Kudlik, S. Benkhof, T. Blochowicz, C. Tschirwitz, and E. A. Rössler, J. Mol. Structure **479**, 201 (1999).
61 S. Kastner, M. Köhler, Y. Goncharov, P. Lunkenheimer, and A. Loidl, J. Non-Cryst. Solids **357**, 510 (2011).
62 A. Rivera and E. A. Rössler, Phys. Rev. B **73**, 212201 (2006).
63 K. Funke, Prog. Solid State Chem. **22**, 111 (1993).
64 J. C. Dyre, J. Appl. Phys. **64**, 2456 (1988).
65 T. B. Schrøder and J. C. Dyre, Phys. Rev. Lett. **101**, 025901 (2008).
66 K. S. Cole and R. H. Cole, J. Chem. Phys. **9**, 341 (1941).
67 A. Puzenko, P. Ben Ishai, and Y. Feldman, Phys. Rev. Lett. **105**, 037601 (2010).
68 A. Faraone, D. V. Wagle, G. A. Baker, E. C. Novak, M. Ohl, D. Reuter, P. Lunkenheimer, A. Loidl, and E. Mamontov, J. Phys. Chem. B **122**, 1261 (2018).
69 D. W. Davidson and R. H. Cole, J. Chem. Phys. **19**, 1484 (1951).
70 H. J. Sillescu, J. Non-Cryst. Solids **243**, 81 (1999).
71 M. D. Ediger, Annu. Rev. Phys. Chem. **51**, 99 (2000).
72 C. Cramer, K. Funke, M. Buscher, A. Happe, T. Saatkamp, and D. Wilmer, Philos. Mag. B **71**, 713 (1995).
73 Y. Wang, W. Chen, Q. Zhao, G. Jin, Z. Xue, Y. Wang, and T. Mu, Phys. Chem. Chem. Phys. **22**, 25760 (2020).
74 C. Krause, J. R. Sangoro, C. Iacob, and F. Kremer, J. Phys. Chem. B **114**, 382 (2010).
75 C. J. Jafta, C. Bridges, L. Haupt, C. Do, P. Sippel, M. J. Cochran, S. Krohns, M. Ohl, A. Loidl, E. Mamontov, P. Lunkenheimer, S. Dai, and X.-G. Sun, ChemSusChem **11**, 3512 (2018).
76 H. Vogel, Z. Phys. **22**, 645 (1921).
77 G. S. Fulcher, J. Am. Ceram. Soc. **8**, 339 (1925).
78 G. Tammann and W. Hesse, Z. Anorg. Allg. Chem. **156**, 245 (1926).
79 C. A. Angell, in *Relaxations in Complex Systems*, edited by K. L. Ngai and G. B. Wright (NRL, Washington, DC, 1985), p. 3.
80 M. D. Ediger, C. A. Angell, and S. R. Nagel, J. Phys. Chem. **100**, 13200 (1996).
81 R. Böhmer, K. L. Ngai, C. A. Angell, and D. J. Plazek, J. Chem. Phys. **99**, 4201 (1993).
82 P. Sippel, P. Lunkenheimer, S. Krohns, E. Thoms, and A. Loidl, Sci. Rep. **5**, 13922 (2015).
83 As $\rho_{dc}(T)$ has negative temperature gradient just as $\langle\tau\rangle(T)$, we use $\rho_{dc}$ instead of $\sigma_{dc}$ in these plots.
84 C. M. Roland, S. H. Bielowka, M. Paluch, and R. Casalini, Rep. Prog. Phys. **68**, 1405 (2005).
85 G. P. Johari and O. Andersson, J. Chem. Phys. **125**, 124501 (2006).
86 S. H. Bielowka, T. Psurek, J. Ziolo, and M. Paluch, Phys. Rev. E **63**, 062301 (2001).
87 As noted in Ref. 84, there is an inconsistent use of the term "Debye-Stokes-Einstein relation" in literature. Sometimes it also denotes a relation between $\eta$ and $\tau$.
88 C. A. Angell, J. Phys. Chem. Solids **49**, 863 (1988)
89 R. Böhmer and C. A. Angell, in *Disorder Effects on Relaxational Processes*, edited by R. Richert and A. Blumen (Springer, Berlin, 1994), p. 11.